\documentclass{PoS}
\usepackage{amsmath}

\title{QCD effects and search for new physics in $t \to b W$ }

\ShortTitle{QCD NP searches in  $t\to bW$ }

\author{\speaker{Svjetlana Fajfer}\\
J. Stefan Institute, Jamova 39, P. O. Box 3000, 1001 Ljubljana, Slovenia \\
       Department of Physics, University of Ljubljana, Jadranska 19, 1000 Ljubljana, Slovenia\\
       \email{svjetlana.fajfer@ijs.si}
                       }
\author{Jure Drobnak \\
     J. Stefan Institute, Jamova 39, P. O. Box 3000, 1001 Ljubljana, Slovenia \\
  \email{jure.drobnak@ijs.si}}
\author{Jernej Kamenik\\
     J. Stefan Institute, Jamova 39, P. O. Box 3000, 1001 Ljubljana, Slovenia \\
         Department of Physics, University of Ljubljana, Jadranska 19, 1000 Ljubljana, Slovenia\\
                       \email{jernej.kamenik@ijs.si}}

\abstract{The most general set of new physics effective operators contributing to the decay of an unpolarized top quark into a bottom quark and a $W$ gauge boson are considered at next-to-leading order in QCD.  We find that the dipole operator contribution to the transverse-plus $W$ helicity fraction $\mathcal F_+$ is  enhanced compared to the leading order result at non-vanishing bottom quark mass. Nonetheless, 
presently the observable most sensitive to new physics contributions is the longitudinal $W$ helicity fraction $\mathcal F_L$.  We also investigate constraints on $t W b$ couplings coming from precision flavor observables, in particular from $\Delta B=1$ and $\Delta B=2$ transitions of $B_q$ mesons. 
We find these to be mostly superior to present direct constraints coming from top decay and single top production measurements at the LHC and Tevatron.}

\FullConference{The 2011 Europhysics Conference on High Energy Physics, EPS-HEP 2011,\\
		July 21-27, 2011\\
		Grenoble, Rh\^one-Alpes, France}

\begin{document}
\section{Introduction}
Top quark physics offers many interesting opportunities to complete our knowledge of Standard Model (SM) physics, as well as to investigate possible appearance of New Physics (NP) effects. 
Recently, both Tevatron experiments D\O \, and CDF  have published results of $W$ helicity measurements in top quark decays~\cite{TeV}:
\begin{align} 
{\cal F}_L^{\rm CDF} \equiv  {\Gamma_L}/{\Gamma} &= 0.88 (13), &
{\cal F}_+^{\rm CDF} \equiv  {\Gamma_+}/{\Gamma} &= - 0.15(9), \nonumber\\
{\cal F}_L^{\rm DO} \equiv  {\Gamma_L}/{\Gamma} &= 0.793(109),  &
{\cal F}_+^{\rm DO} \equiv  {\Gamma_+}/{\Gamma} &= -0.002(05).\label{e1}
\end{align}
In the SM, simple helicity considerations show that ${\cal F}_+$  vanishes at the Born term level in the limit of vanishing mass of the $b$ quark. Conversely, a non-vanishing transverse-plus rate within the SM appears due to the nonzero $b$ quark mass as well as due to radiative QCD corrections in the form of real gluon emission,  and amounts to only a $0.1\%$  effect. Another possibility to obtain nonzero ${\cal F}_+$  comes from contributions beyond the SM.  Also in this case however QCD effects can play an important role when relating possible NP effects in $tWb$ interactions to the  top quark decay rate and helicity fraction measurements.
On the other hand, the same $tWb$ interactions contribute also in $B$ physics, since Flavor Changing Neutral Current (FCNC) processes involving $b$ quarks receive, within the SM, dominant contributions from loops involving a top quark and a $W$ boson. Interesting effects of the anomalous $tWb$  interactions can thus appear in observables related to $B_{d,s }  - \bar B_{d,s}$ oscillations  as well as rare $b\to s (\gamma, \ell^+ \ell^- , \nu\bar\nu)$ transitions.
 \section{Framework} 
Our studies are based on a general effective Lagrangian for the $tWb$ interaction, which appears in presence of NP heavy degrees of freedom, integrated out at a scale above the top quark mass.  The NP effective operators include a modification of the SM charged current by left (right) - handed  currents, as well as dipole operators. In general the  complete $t \to bW$ decay width can be written
  as a sum of decay widths distinguished by different  helicities of the $W$ boson
 \begin{eqnarray}  
 \Gamma(t \to b W) = \sum_i \Gamma_i\,, \quad \mathcal F_i \equiv {\Gamma_i}/{\Gamma}\,, \quad {\rm thus} \quad\sum_i \mathcal F_i = 1\,.
 \label{e2}
 \end{eqnarray}
In~\cite{JD1}, we have computed all ${\cal O} (\alpha_s)$ corrections to the polarized rates $\Gamma_{i}$, where $i=L,\pm$ (given in Fig. 2 of \cite{JD1}) in presence of effective NP contributions.  Since in the $m_b = 0$ limit the Leading Order (LO) SM contribution to $\Gamma_{+}$ vanishes, we use the full $m_b$ dependence of the LO rates, but we neglect all ${\cal O} (\alpha_s m_b/m_t)$ contributions. We observe that the modification of NP effects is substantial when going to Next-to-Leading Order (NLO) in
QCD, however in $\mathcal F_+$ they remain at most at the $1-2$ per-mille level. In particular, a nonstandard value of the left-handed $tWb$ current coupling does not affect the
different $W$ helicity branching fractions at all. On the other hand nonzero $tWb$ interactions involving the right-handed $b$ quark could significantly affect ${\cal F}_+$, but are severely constrained by indirect bounds from $B$ physics~\cite{JD2,JD3}.  Our analysis also includes NLO QCD effects in ${\cal F}_L$. Considering a single real NP contribution at the time, we find that QCD corrections decrease NP effects in ${\cal F}_L$ by approximately $1\%$  in all cases. 

We have continued our study of anomalous $tWb$ interactions in $\Delta B=2$ and $\Delta B=1$ transitions within the effective theory accompanied by the principle of Minimal Flavor Violation (MFV)~\cite{JD2,JD3}.   Our operator basis consists of all dimension-six operators that generate charged current quark interactions with the $W$, but do not induce FCNCs at the tree-level. Since we restrict our discussion to MFV scenarios, we first identify four relevant quark bilinears with distinct transformation properties under the
SM gauge symmetry group, resulting in seven effective flavored operators all involving the $tWb$ vertex as described in~\cite{JD2}.  After establishing our operator basis relevant for $B_q - \bar B_q$, $b \to s \gamma$ and $b \to s l^+ l^-$,  we perform matching of  our effective theory to the low energy theory, by integrating out the top quark and electroweak gauge boson contributions at leading order QCD.  With this at hand, we can calculate the resulting contributions to $B_q - \bar B_q$ mixing amplitudes as well as $B\to X_s\gamma$ and $B\to X_s \ell^+ \ell^-$  decay rates.  In our calculation we neglect the masses of the strange quark and light leptons.  
The bounds on real parts of Wilson coefficients come from the most precise measurements of $\Delta m_{s,d}$, $BR( B \to X_s \gamma)$ and $BR( B \to X_s  \ell^+ \ell^-)$ at low $q^2$. 
We observe that the  $\Delta B=2$ mixing amplitudes can receive sizable contributions allowed by $\Delta B=1$ observables and direct constraints, and that within this framework significant new sources of CP violation can appear. In particular, the imaginary parts of some Wilson coefficients can be determined from the experimentally measured mixing phases of $B_{d,s}$ mesons~\cite{JD2} as well as the CP asymmetry in  the $B \to X_s \gamma$ decay~\cite{JD3}. Collecting all bounds from $\Delta B=1$ and $\Delta B=2$  processes we combine and compare them with direct constraints coming from the $t \to b W$ decay and single top production measurements. Allowing only a single non-zero NP operator contribution at the time, we find the combined indirect constraints on most operators to be more stringent than direct bounds. Having more than one observable at disposal one can also consider pairs of operators contributing simultaneously  and obtain  allowed regions in the corresponding planes as shown in the second and third plot of Fig. 4. in~\cite{JD3}.   
\section{Conclusions}   
We have investigated contributions of anomalous $tW b$ interactions  to the decay of a top quark to a $W$ gauge boson and a $b$ quark. We have analyzed the impact of QCD corrections on the most general parametrization of such effective NP contributions in this process. QCD corrections were found to be small for the ${\cal F}_L$ helicity amplitude reaching at most the $1\%$ level. While larger effects were found in $\mathcal F_+$, any sizable NP contributions to this observable are subject to severe indirect $B$ physics constraints. In term we have determined such indirect bounds on the real and imaginary parts of the anomalous $tWb$ couplings. For most of the considered effective operators, the indirect  bounds are at present much stronger than the direct constraints coming from the $t \to bW $ helicity fractions, angular asymmetries and single top production measurements at the Tevatron and the LHC. In particular, we were able for the first time to constrain the imaginary parts of most of the anomalous couplings. Taking into account these bounds, we have predicted the presently allowed effects of the anomalous $tWb$ interactions on the $B_s \to \mu^+ \mu^-$  decay rate, the forward-backward asymmetry in $B \to K^* \ell^+ \ell^-$, as well as the branching ratios of $B\to K^{(*)} \nu \bar \nu$  decays. In the future, more precise measurements of the $t \to b W$ helicity fractions as well as $\Delta B=1$ and $\Delta B=2$ rare $B$ physics processes could further constrain anomalous $tWb$ couplings.

\end{document}